\documentclass{ws-procs975x65}

\begin{document}



\title{Impact of dark matter decays and annihilations on structure formation}

\author{EMANUELE RIPAMONTI}

\address{Kapteyn Astronomical Institute, University of Groningen, Postbus 800, 9700 AV,\\ Groningen, The Netherlands, \email{ripa@astro.rug.nl}}

\author{MICHELA MAPELLI}

\address{SISSA/ISAS,
Via Beirut 2-4, 
Trieste I-34014, Italy 
, \email{mapelli@sissa.it}}


\begin{abstract}
We derived the influence of dark matter (DM) decays and annihilations on
structure formation. The energy deposited by DM decays and annihilations
into metal free halos both increases the gas temperature and enhances the formation of molecules. Within the primordial halos the temperature increase generally dominates over the molecular cooling, slightly delaying the collapse. In fact, the critical mass for collapse is generally higher than in the unperturbed case, when we consider the energy input from DM. 
In presence of DM decays and/or annihilations the fraction of baryons inside collapsed metal free halos should be slightly less ($\approx{}0.4$) than the expected cosmological value.
\end{abstract}

\bodymatter

\section{Introduction}\label{intro}

The formation of the first luminous objects is heavily influenced both by the chemical abundance of coolants and by any source of heating. In particular, higher temperatures can prevent the collapse of halos, whereas a greater abundance of H$_2$ and HD molecules (which are the main coolants of the metal free Universe) enhances the cooling of the gas, favouring the collapse. In principle, reionization sources can both enhance the abundance of molecules, since free electrons act as catalysts of H$_2$ and HD, and increase the temperature of the gas. 

\section{Heating and molecular abundance enhancement from DM}
DM decays and annihilations can be sources of heating and partial early reionization\cite{RMF06}. Thus, they are also expected to affect the abundance of molecules. 
Fig.~1.a shows the temperature and the fractional abundance of free electrons, H$_2$ and HD in  the intergalactic medium (IGM) as a function of redshift in the unperturbed case (solid line) and if we switch on different models\footnote{The effects of DM decays and annihilation on heating and ionization have been calculated taking into account the fraction of energy which is effectively absorbed by the gas\cite{RMF06}.} of DM decays and annihilations (dashed line; from top to bottom: sterile neutrino decays, light dark matter (LDM) decays and LDM annihilations). DM decays and annihilations both heat the gas\cite{RMF06} and increase the abundance of free electrons, which enhances the formation of H$_2$ and HD.

To infer what is the net effect of DM decays and annihilations on structure formation, we have to follow the evolution of a large grid of metal free halos. For this purpose, we used a one-dimensional Lagrangian code\cite{R06}, which simulates the gravitational and hydro-dynamical evolution of the gas, accounting for the evolution of 12 chemical species, for the cooling/heating effects and for the gravitational influence of the DM halo. We included into the code the effects of DM decays and annihilations\cite{RMF06}.
Fig.~1.b shows the behaviour of gas density, temperature, ionization fraction and H$_2$ fractional abundances within a simulated halo of $6\times{}10^5\,{}M_\odot{}$ virializing at redshift 12. Also within the halo DM decays and annihilations enhance both the temperature and the molecular abundance. However, from the plot of the density, we can see that for the considered DM models the collapse is delayed by decays/annihilations, and even prevented in the case of LDM decays\cite{RMF2}.

\def\figsubcap#1{\par\noindent\centering\footnotesize(#1)}

\begin{figure}\label{fig:fig1}%
\begin{center}
 \parbox{2.15in}{\epsfig{figure=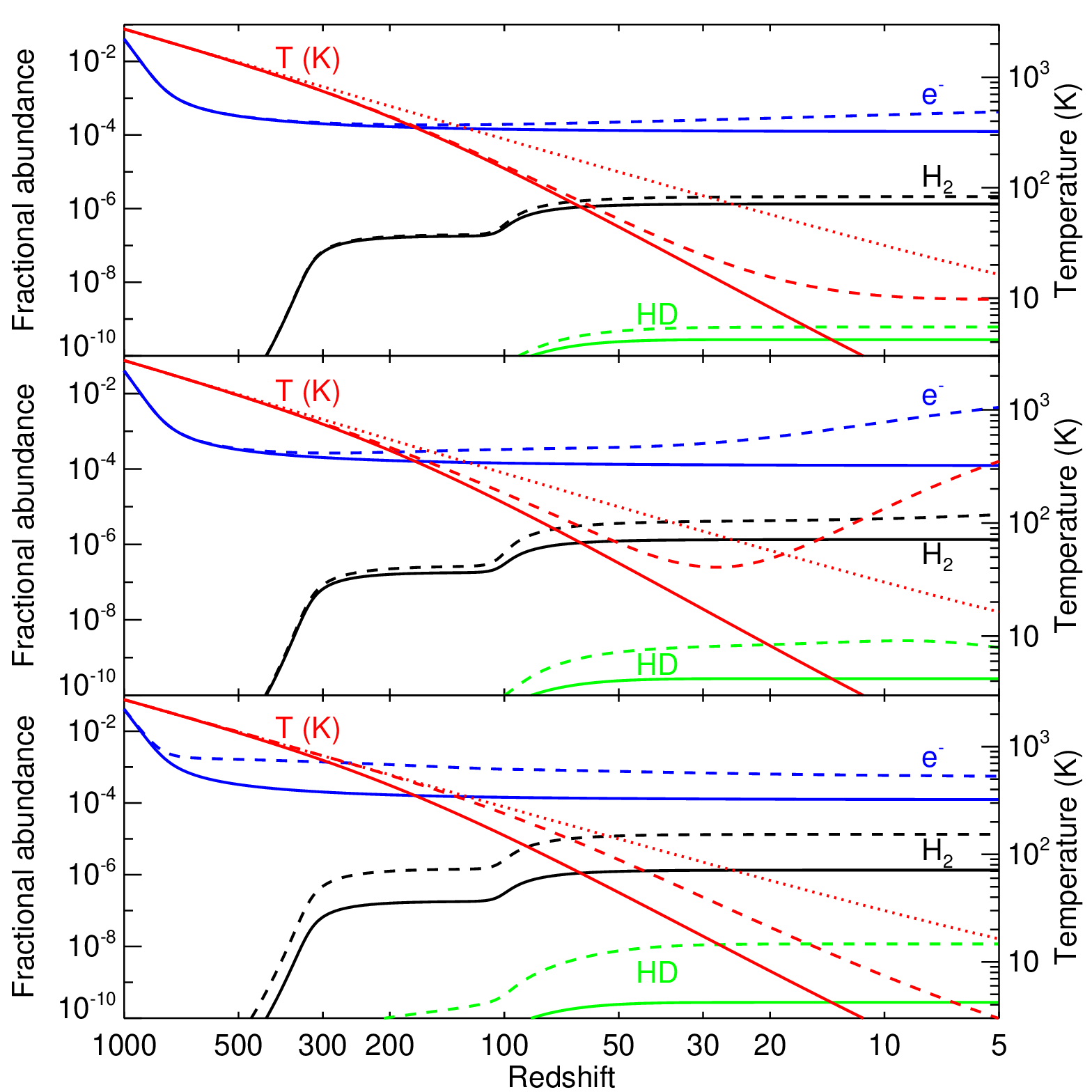,width=2.1in}
 \figsubcap{a}}
 \hspace*{4pt}
 \parbox{2.15in}{\epsfig{figure=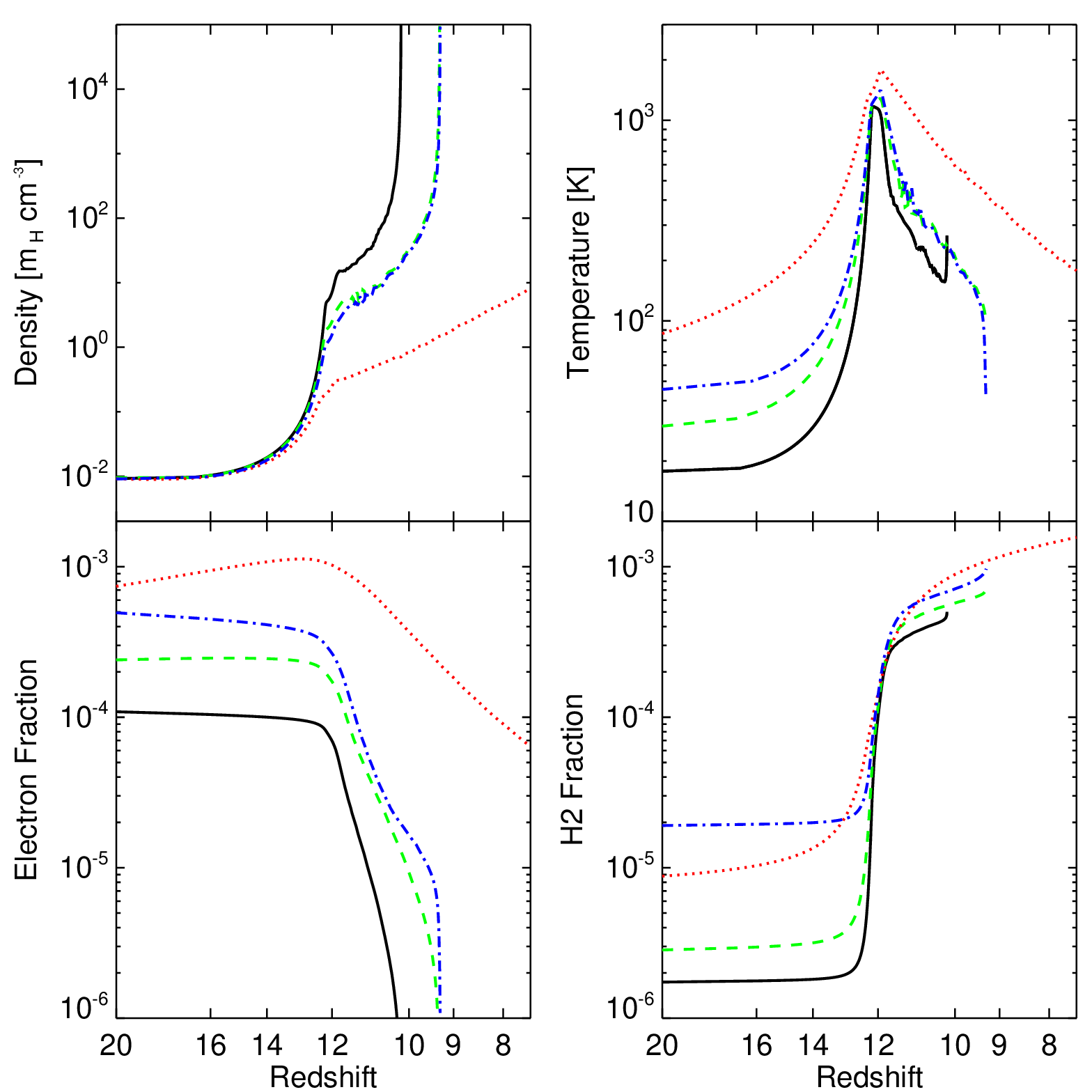,width=2.1in}
 \figsubcap{b}}
\caption{(a) Effects of decaying/annihilating DM on the IGM evolution. Left axis: fractional abundances of free electrons ($e^-$), H$_2$ and HD as a function of the redshift. Right axis: matter temperature as a function of the redshift. Top panel: Effect of decaying sterile neutrinos of mass 25 keV (dashed line).  Central panel:  decaying LDM of mass 10 MeV (dashed line).  Bottom panel: annihilating LDM of mass 1 MeV (dashed line). The dotted line is the CMB temperature and the solid line represents the thermal and chemical evolution without DM decays/annihilations. (b) Evolution of the central region of a
$6\times{}10^5\,{}M_\odot{}$ halo virializing at
$z_{vir}=12$.  From left to right and from top to bottom: density,
temperature, electron abundance and H$_2$ abundance as function of
redshift. The solid line represents the unperturbed case. The dashed, dot-dashed and dotted lines
account for the contribution of 25-keV sterile neutrino decays, 1-MeV
LDM annihilations and 10-MeV LDM decays, respectively.} 
\end{center}
\end{figure}

\section{Critical mass and gas content in metal free halos}
\begin{figure}[b]\label{fig:fig2}%
\begin{center}
 \parbox{2.15in}{\epsfig{figure=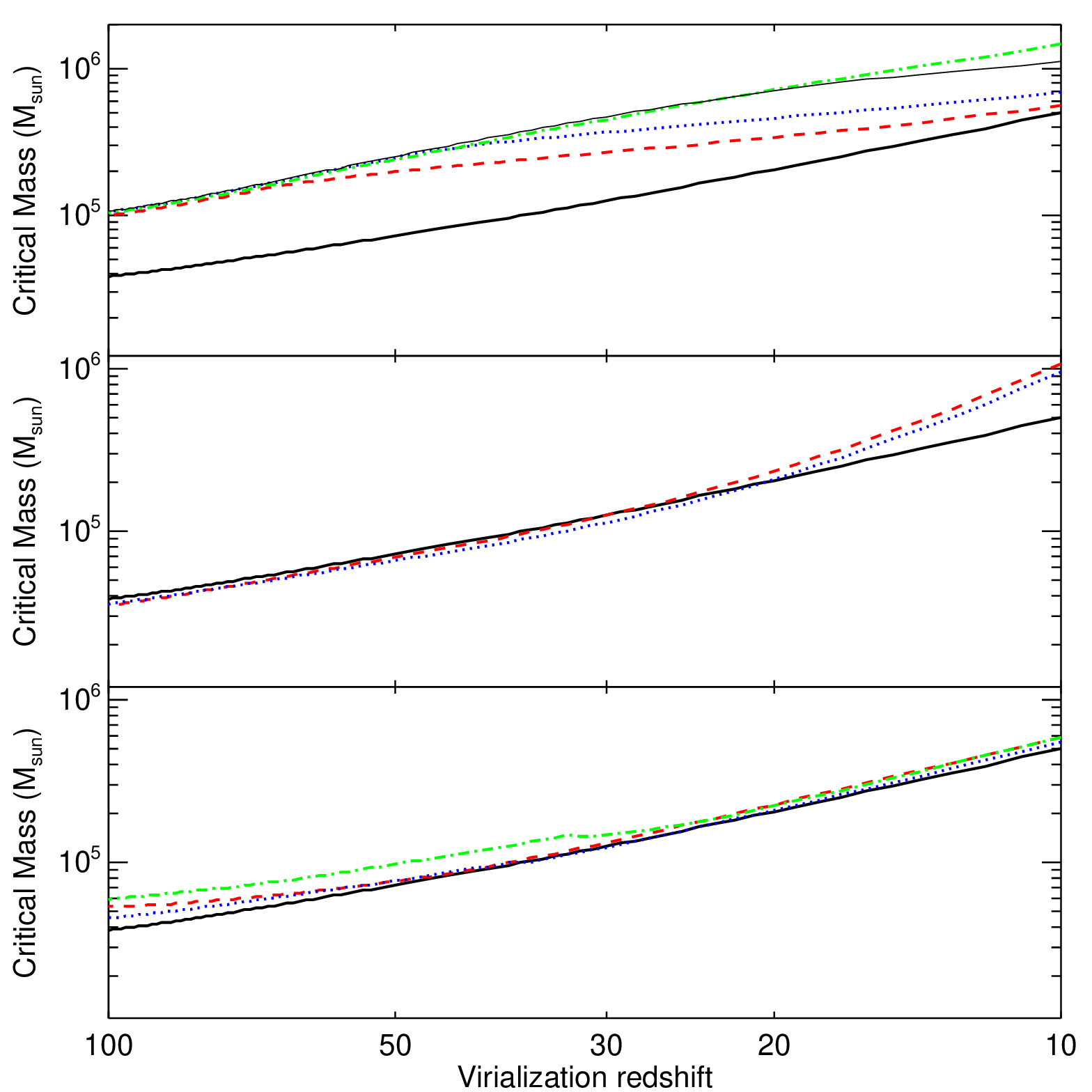,width=2.1in}
 \figsubcap{a}}
 \hspace*{4pt}
 \parbox{2.15in}{\epsfig{figure=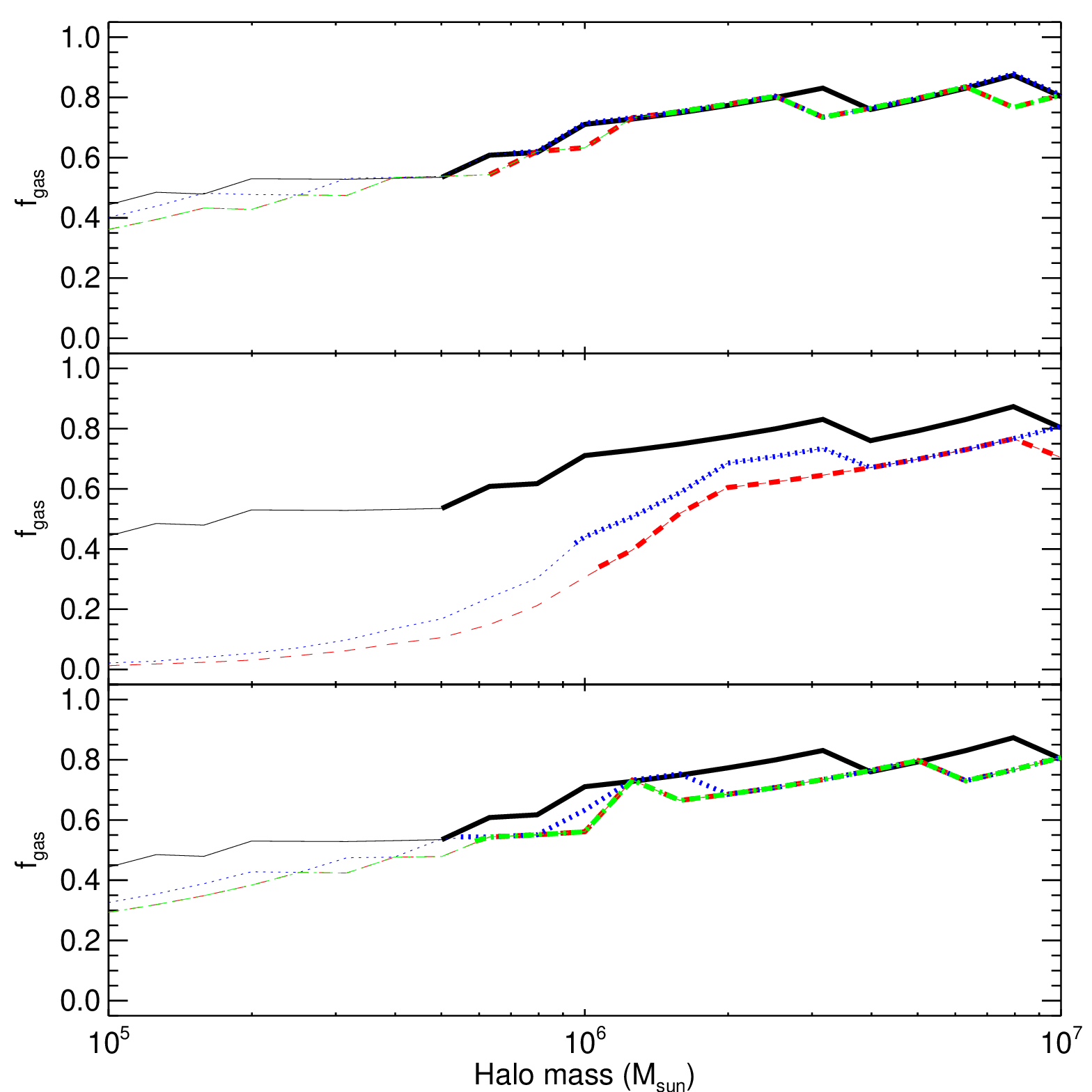,width=2.1in}
 \figsubcap{b}}
 \caption{Top panel: Decaying sterile neutrinos of 4 (dot-dashed line), 15 (dotted) and 25 keV (dashed). Central: Decaying LDM of 3 (dotted line) and 10 MeV (dashed). Bottom: Annihilating LDM of 1 (dot-dashed line), 3 (dotted) and 10 MeV (dashed). The solid lines represent the unperturbed case.
(a) $m_{crit}$ as a function of the
virialization redshift $z_{vir}$. (b) Halo baryonic mass fraction as a
function of the halo mass for a
fixed virialization redshift ($z_{vir}=10$). Thick (thin) lines indicate that the
halo mass is larger (smaller) than $m_{crit}$. }
\end{center}
\end{figure}
In order to quantify these considerations, Fig.~2.a shows the critical mass $m_{crit}$ (i.e. the minimum halo mass for collapse at a given redshift) as a function of the virialization redshift, in presence (from top to bottom) of sterile neutrino decays, LDM decays and LDM annihilations.
$m_{crit}$ is generally increased by DM decays and annihilations, confirming that these tend to delay the collapse of metal free halos. However, the difference with respect to the unperturbed case is less than a factor 2-4 (depending on the redshift and on the model), indicating that the effect of DM decays and annihilations on structure formation is quite negligible\cite{RMF2}.

On the other hand, DM decays and annihilations might have important effects on the baryonic content of such small metal free halos. Fig.~2.b shows $f_{\rm gas}$, i.e. the ratio between the amount of gas which is contained within the virial radius of the simulated halos and the mass of gas which we should expect from cosmological parameters, as a function of the halo mass. The solid line shows the unperturbed case, while the dashed, dotted and dot-dashed lines indicate various models of DM decays and annihilations. $f_{gas}$ is always smaller if we switch on the contribution by DM decays and annihilations. In particular, in the case of LDM decays, $f_{gas}$ drops to $\sim{}0.4$ (if we consider only collapsed halos) or even to $\sim{}0.02$ (if we include smaller, non-collapsed halos). In conclusion, this means that, in presence of LDM decays, small metal free halos can still collapse ($m_{crit}$ being almost unperturbed); but their baryonic content is smaller than in larger halos, and, as a consequence, they can form a smaller total mass of stars than expected.

\section*{Acknowledgments}
ER acknowledges support from NWO grant 436016. MM acknowledges the organizers of the Eleventh Marcel Grossmann meeting for the MGF grant. The authors thank P. L. Biermann for inviting them to the meeting. 


\vfill

\end{document}